# POSSIBILITY OF CONTROL OF THE GRAVITATIONAL MASS BY MEANS OF EXTRA-LOW FREQUENCIES RADIATION

Fran De Aquino[*]

Physics Department, Maranhao State University, S.Luis/ MA, Brazil.

According to the *weak* form of Einstein's general relativity *equivalence principle*, the gravitational and inertial masses are equivalent. However recent calculations have revealed that they are correlated by an adimensional factor, which is equal to one in absence of radiation only. We have built an experimental system to check this unexpected theoretical result. It verifies the effects of the extra-low frequency (ELF) radiation on the *gravitational mass* of a body. We show that there is a direct correlation between the radiation absorbed by the body and its gravitational mass, independently of the inertial mass.

## Introduction

The physical property of mass has two distinct aspects, *gravitational mass* $m_g$ and *inertial mass* $m_i$. Gravitational mass produces and responds to gravitational fields. It supplies the mass factors in Newton's famous inverse-square law of gravity ($F_{12} = G m_{g1} m_{g2} / r_{12}^2$). Inertial mass is the mass factor in Newton's 2nd Law of Motion (F=$m_i$a). One of the deep mysteries of physics is the correlation between these two aspects of mass. Several experiments[1-6] have been carried out since the century XIX to try to verify the correlation between gravitational mass $m_g$ and inertial mass $m_i$.

In a recent paper[7] we have shown that the *gravitational mass* and the *inertial mass* are correlated by an adimensional factor, which depends on the incident radiation upon the particle. It was shown that only in the absence of electromagnetic radiation this factor becomes equal to 1 and that, in specific electromagnetic conditions, it can be reduced, nullified or made negative. This means that there is the possibility of control of the gravitational mass by means of the incident radiation.

The general expression of correlation between gravitational mass $m_g$ and inertial mass $m_i$, is given by

$$m_g = m_i - 2\left\{\sqrt{1+\left\{\frac{U}{m_i c^2}\sqrt{\frac{\varepsilon_r \mu_r}{2}\left(\sqrt{1+(\sigma/\omega\varepsilon)^2}+1\right)}\right\}^2} - 1\right\}m_i \quad (1)$$

The electromagnetic characteristics, $\varepsilon$, $\mu$ and $\sigma$ do not refer to the particle, but to the outside medium around the particle in which the incident radiation is propagating. For an *atom* inside a body, the incident radiation on this atom will be propagating inside the body, and consequently, $\sigma = \sigma_{body}$, $\varepsilon = \varepsilon_{body}$, $\mu = \mu_{body}$. So, if $\omega \ll \sigma_{body}/\varepsilon_{body}$, equation above reduces to

$$m_g = m_a - 2\left\{\sqrt{1+\left\{\frac{U}{m_a c^2}\sqrt{\frac{c^2 \mu_{body} \sigma_{body}}{4\pi f}}\right\}^2} - 1\right\}m_a \quad (2)$$

where $m_a$ is the *inertial* mass of the atom.

Thus we see that, *atoms* (or *molecules*) can have their *gravitational masses* strongly reduced by means of extra-low frequency (ELF) radiation.

We have built a system to verify the effects of the ELF radiation on the gravitational mass of a body. In this work we present the experimental set-up and the results obtained.

## Experimental

Let us consider the apparatus in figure 1. The Transformer has the following characteristics:

- Frequency : 60 Hz

---

[*] Permanent Address: R.Silva Jardim, 521-centro, 65020-560 S. Luis/MA, Brazil.( E-mail: deaquino@uema.br ).

- Power : 11.5kVA
- Number of turns of coil : $n_1 = 12$, $n_2 = 2$
- Coil 1 : copper wire 6 AWG
- Coil 2 : ½ inch diameter copper rod (with insulation paint).
- Core area:502.4cm$^2$ ; $\phi$=10 inch (Steel)
- Maximum input voltage : $V_1^{max}$ = 220 V
- Input impedance : $Z_1 = 4.2\ \Omega$
- Output impedance : $Z_2 < 1m\ \Omega$ ( ELF antenna impedance : 116 m$\Omega$ )
- Maximum output voltage with coupled antenna : 34.8V
- Maximum output current with coupled antenna : 300 A

In the system-G the *annealed pure iron* has an electric conductivity $\sigma_i$ =1.03×10$^7$S/m, magnetic permeability $\mu_i = 25000\mu_0$ [8], thickness 0.6 mm ( to absorb the ELF radiation produced by the antenna). The *iron powder* which encapsulates the ELF antenna has $\sigma_p \approx$10 S/m ; $\mu_p \approx 75\mu_0$ [9]. The antenna physical length is $z_0$ = 12 m, see Fig.1c. The power radiated by the antenna can be calculated by the well-known *general* expression, for $z_0 \ll \lambda$ :

$$P = (I_0\ \omega z_0)^2 / 3\pi\varepsilon v^3 \{[1+ (\sigma/\omega\varepsilon)^2]^{½} +1\} \quad (3)$$

where $I_0$ is the antenna current amplitude ; $\omega = 2\pi f$ ; $f$ =60Hz ; $\varepsilon = \varepsilon_p$ ; $\sigma = \sigma_p$ and $v$ is the wave phase velocity in the *iron powder* ( given by Equation1.02 , in reference [1] ). The radiation efficiency $e = P / P+P_{ohmic}$ is nearly 100%.

Each atom of the annealed iron toroid absorbs an ELFenergy $U=\eta P_a /f$, where $\eta$ is a particle-dependent absorption coefficient (the maxima $\eta$ values occurs, as we know, for the frequencies of the atom's *absorption spectrum* ) and $P_a$ is the incident radiation power on the atom ; $P_a=DS_a$ where $S_a$ is the atom's *geometric* cross section and $D=P/S$ the radiation power density on the iron atom ( $P$ is the power radiated by the antenna and $S$ is the area of the annealed iron toroid ($S$ = 0.374 m$^2$, see Fig.1b) . So, we can write :

$$U = \eta S_a(I_0 z_0)^2 \omega / 3S\varepsilon_i v^3 \{[1+(\sigma_i/\omega\varepsilon_i)^2]^{½}+1\} \quad (4)$$

Consequently, according to Eq.(1) , for $\omega \ll \sigma_i/\varepsilon_i$ , the gravitational masses of these iron atoms, under these conditions, will be given by :

$$m_g = m_a - 2\{[1+4.4\times10^{-9}\ I_0^4]^{½} - 1\}m_a \quad (5)$$

Equation above shows that the *gravitational masses* ($m_g$) of the atoms of the annealed pure iron toroid can be *nullified* for $I_0 = 129.83$A. Above this critical value the gravitational masses becomes negatives (antigravity).

### Results and Discussion

Figure 2 presents the results of $m_g$ calculated by means of Eq.5, plotted as a function of current $I_0$ , for $\mu_i = 25000\mu_0$ ; $\sigma_i = 1.03\times10^7$ S/m ; $\sigma_p \approx$10 S/m ; $\mu_p \approx 75\mu_0$ ; $z_0$ = 12 m.

The experimental results obtained ( see Table1) are plotted on said figure to be compared with those supplied by the theory.

It is important to note that, in practice, when $I_0$ = 130.01A the gravitational mass of system-G reduces to 5.80kg ; exactly equal to the mass of the *steel toroid* (see fig.1). This occurs due to gravitational mass of the annealed pure iron toroid to become null when $I_0$ =130.01A (Exactly as predicted by the theory. i.e., the gravitational masses of the atoms of the annealed iron toroid become *null* for $I_0 = 129.83$A).

Under these circumstances, the toroid doesn't interact gravitationally with the Universe, and consequently , there can be no gravitational interaction between the matter inside the toroid and the rest of the Universe. Therefore, the gravitational mass of system-G reduces to the mass of the *steel toroid* , which is outside the annealed iron toroid.

### Conclusion

This experiment ( carried out by the author on January 27, 2000)

provides a strong evidence that the general expression of correlation between gravitational mass and inertial mass (Eq.1) is true. So, we can easily conclude that the gravitational forces can be reduced, nullified and inverted by means of electromagnetic radiation .

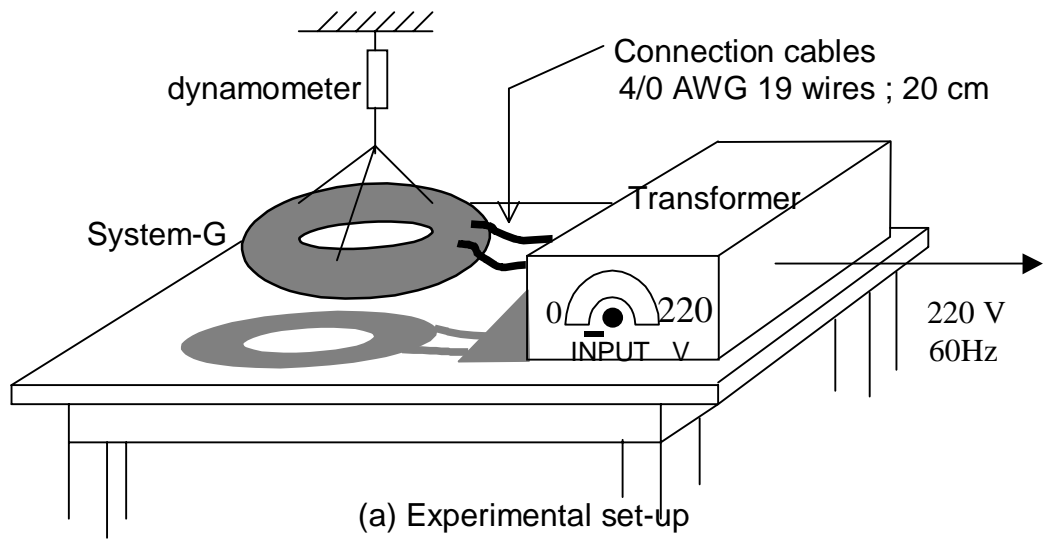

(a) Experimental set-up

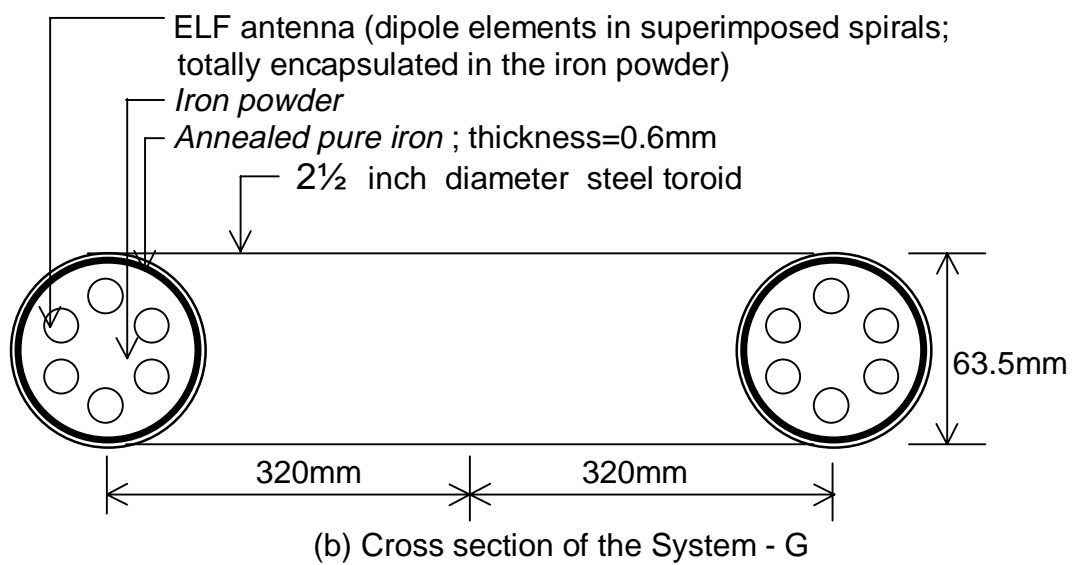

(b) Cross section of the System - G

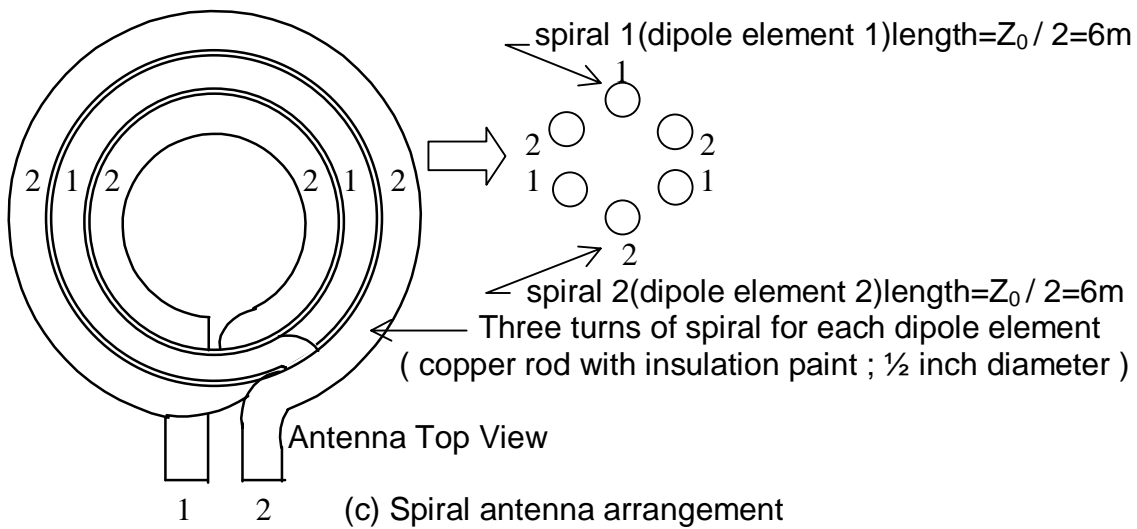

(c) Spiral antenna arrangement

Fig. 1 – Schematic View of the Experimental Apparatus

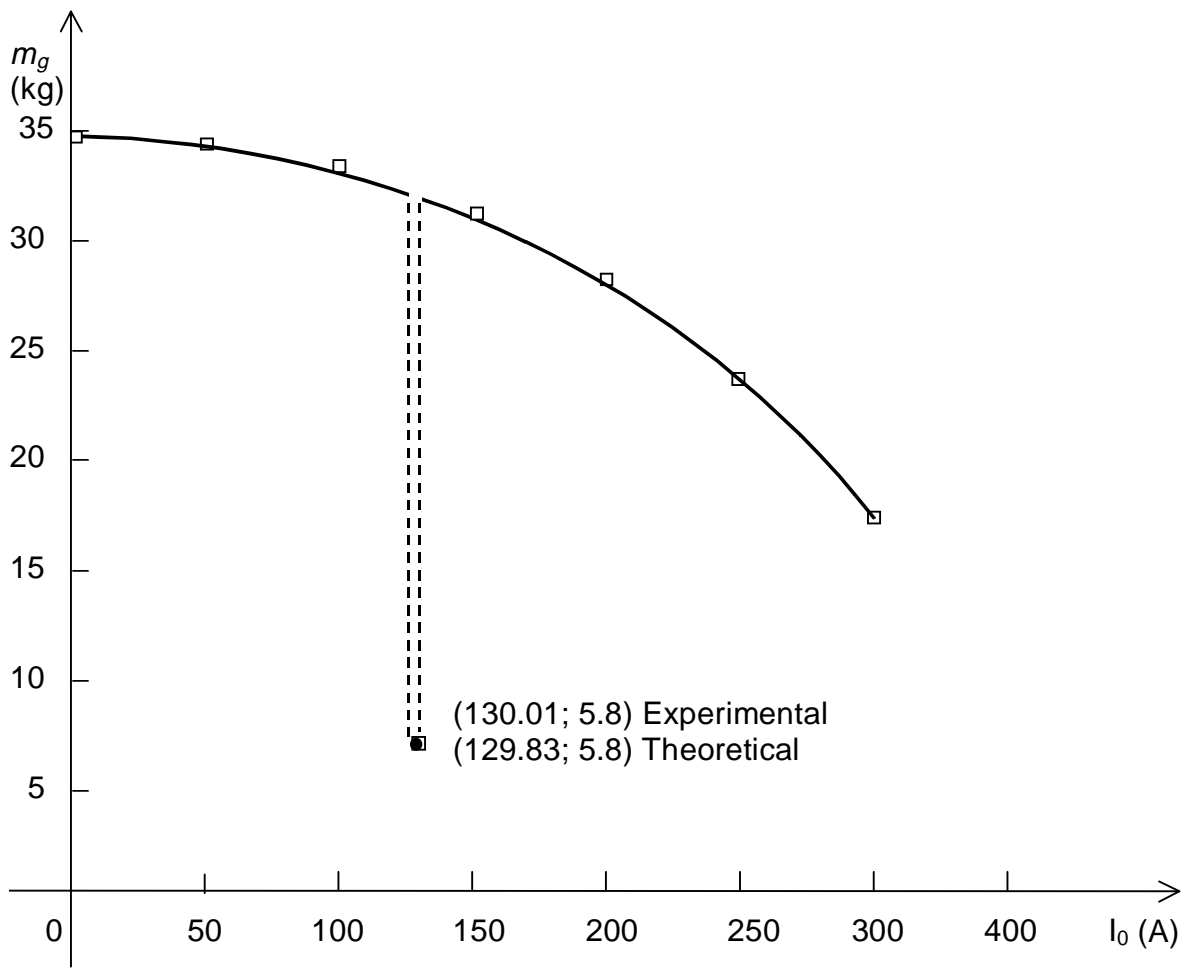

Fig.2 – Comparison between experimental data(□) and theory (solid line).

| $I_0$ (A) | $m_g$ (kg) | |
| --- | --- | --- |
| | Theory | Exper. |
| 0 | 34.85 | 34.85 |
| 50 | 34.80 | 34.83 |
| 100 | 34.17 | 34.26 |
| 130.01 | 5.80 | 5.80 |
| 150 | 32.14 | 32.25 |
| 200 | 28.61 | 28.68 |
| 250 | 23.75 | 23.80 |
| 300 | 17.68 | 17.69 |

Table 1